\documentclass[%
 reprint,
 amsmath,amssymb,
 aps,
]{revtex4-1}
\usepackage[latin1]{inputenc}
\usepackage{mathtools}
\usepackage{slashed}
\usepackage{physics}
\usepackage{array}
\usepackage{amsmath}
\usepackage[dvipsnames]{xcolor}
\usepackage{amsfonts}
\usepackage{amssymb}
\usepackage{relsize}
\usepackage{amssymb,epsfig,subfigure}
\usepackage{hyperref}
\usepackage{comment}
\usepackage{tabularx}
\usepackage{bm}
\usepackage{euscript}
\usepackage{graphicx}
\usepackage{mathrsfs}
\usepackage[scr=boondox]{mathalfa}
\usepackage{color}
\usepackage [english]{babel}
\usepackage [autostyle, english = american]{csquotes}
\MakeOuterQuote{"}
\usepackage{amsfonts}
\usepackage{exscale}
\usepackage{amsbsy}
\usepackage{textcomp}
\usepackage{comment}
\usepackage{slashed}
\usepackage{tabularx}
\usepackage{euscript}
\usepackage{graphicx}
\usepackage{color}
\usepackage{amsfonts}
\usepackage{exscale}
\usepackage{amsbsy}
\usepackage{subfigure}
\usepackage{textcomp}
\usepackage{comment}
\usepackage{hyperref}
\usepackage{bm}
\usepackage{wrapfig}
\usepackage[singlelinecheck=false,justification=justified]{caption}
\usepackage{ulem}

\newcommand{\beq}{\begin{eqnarray}}
\newcommand{\eeq}{\end{eqnarray}}
\begin{document}

\def\ppnumber{\vbox{\baselineskip14pt
}}

\def\ppdate{
} \date{\today}

\title{\bf Anomalous flux state in higher-order topological superconductors}
\author{Yizhi You}
\affiliation{Department of Physics, Northeastern University, MA, 02115 USA}

\begin{abstract}
We investigate the anomalous flux state of interacting higher-order topological superconductors (HOTSC) protected by rotation symmetries.
By introducing a $\pi$ superconducting flux in 2D HOTSC, we demonstrate the existence of a robust zero mode trapped at the flux center. Remarkably, the rotation symmetry and fermion parity display projective representation inside the $\pi$ flux with $N=2$ supersymmetry algebra. A similar gapless flux pattern also exists in 3D HOTSCs, the flux lines of which carry anomalous helical modes that cannot be realized on purely one-dimensional lattice models. Notably, these exotic phenomena can be manifested in a 2D frustrated quantum magnet whose low energy excitation characterizes emergent Majoranas with HOTSC band structure and the $Z_2$ flux exhibiting supersymmetries.
\end{abstract}

\maketitle

\bigskip
\newpage

\section{Introduction}
Higher-order topological superconductors (HOTSC) are novel forms of gapped quantum matter that host gappable surfaces but gapless corners or hinges in-between\cite{benalcazar2017quantized,schindler2018higher}. Since their initial discovery, HOTSC and their descendants have been discussed extensively and have become an active area of theoretical and experiential research. Recent progress has included topological classifications\cite{benalcazar2017electric,song2017d,langbehn2017reflection,khalaf2018higher,benalcazar2019quantization,you2021multipolar}, topological field theories\cite{you2021multipolar, may2021crystalline}, and experimental realization of various classes of HOTSC\cite{noh2018topological,serra2018observation,peterson2018quantized,imhof2018topolectrical,schindler2018higherb,xue2019acoustic,zhang2019second,ni2019observation,noguchi2021evidence,aggarwal2021evidence}. 

Despite the rapid progress in the understanding of higher-order topological superconductors from a band structure perspective~\cite{isobe2015theory,huang2017building,song2017interaction,song2017topological,you2018higher,rasmussen2018intrinsically,rasmussen2018classification,thorngren2018gauging,benalcazar2018quantization,zhang2019construction,tiwari2019unhinging,you2018higher,jiang2019generalized}, experimentally accessible fingerprints for observing HOTSC still remain challenging in strongly correlated systems. Notably, the observation of gapless Majorana modes at the corners or hinges does not fully guarantee that the bulk is a higher-order topological superconductor (HOTSC), as some of these gapless modes can potentially be annihilated via surface gap closing, even if the bulk spectrum remains gapped\cite{you2019multipolar,tiwari2019unhinging}.
Alternatively, some higher-order topological superconductors (HOTSCs) can exhibit fully gappable boundaries, including corners and hinges~\cite{tiwari2019unhinging}, while still exhibiting a non-trivial entanglement structure that distinguishes them from trivial superconductors. Previous works have established that higher-order topological insulators and superconductors can be probed by their geometric responses, such as the creation of lattice defects like dislocations or disclinations. By creating these defects, one can observe Majorana zero modes inside the disclination point in two dimensions or chiral fermion modes localized at the dislocation/disclination lines in three dimensions\cite{liu2019shift,you2018highertitus,teo2012majorana,li2020fractional,may2021crystalline,queiroz2019partial,zhang2022bulk,schindler2022topological}. In contrast to these approaches, which are primarily based on the non-interacting limit, we aim to identify the universal fingerprints and topological responses specifically for strongly interacting higher-order topological superconductor (HOTSC) phases.

In this work, we seek to unravel the nature of superconducting $\pi$ flux and their corresponding topological response features in 2D and 3D HOTSC. First, we begin with the 2D HOTSC model on a square lattice proposed in Eq.~\cite{wang2018weak} protected by $C_4$ symmetry. We demonstrate that creating a $\pi$ superconducting flux engenders a projective symmetry between $C_4$ and fermion parity $P$, so the resultant flux state contains a protected two-fold degeneracy. Remarkably, this projective symmetry within the flux uniquely generates the $N=2$ supersymmetry (SUSY) algebra in quantum mechanics\cite{hsieh2016all}.

Motivated by these observations, we extend our horizon into frustrated spin systems whose emergent quasiparticles and flux excitations exactly reproduce the topological feature of HOTSC\cite{dwivedi2018majorana}. Namely, we construct a bosonic spin-$\frac{3}{2}$ model whose low-energy excitations contain emergent Majoranas coupling with a $Z_2$ gauge field. The Majorana forms a superconducting band akin to the 2D HOTSC, while the $Z_2$ gauge flux excitation carries $N=2$ SUSY structure.

In section\ref{sec:3d}, we examine the role of superconducting $\pi$ flux in 3D HOTSC with $C_4^T$ symmetry. One of our main findings is that the flux lines inside the HOTSC trap 1D helical Majorana modes with an intrinsic quantum anomaly. In particular, the gapless modes inside the flux line are anomalous in the sense that the $C_4^T$ symmetry will inevitably be broken if we gauge the fermion parity inside the flux line. 
Under this observation, the helical Majorana modes inside the flux exhibit global anomaly, signaling the impossibility of realizing them on an isolated one-dimensional lattice model.
Notably, such quantum anomaly manifested by `conflict of the symmetries' had been widely observed in the surface theory of symmetry-protected topological phase\cite{ryu2015interacting,cho2017relationship}.

Our result provides a new route to detect HOTSC in numerical simulations via flux responses. The projective symmetry in the 2D HOTSC flux state can be detected from the shift of the entanglement spectrum upon flux insertion. Suppose one creates a rotational symmetric cut of the ground state wave function after $\pi$ flux insertion; the entanglement spectrum will display a robust two-fold degeneracy in all spectrum levels. This degeneracy persists even for small system sizes, where finite-size effects are inevitable. More precisely, the whole entanglement Hamiltonian develops a projective symmetry under rotation and fermion parity after flux insertion, which results in a degenerate spectrum in the entanglement Hamiltonian for both the ground state and highly excited states. Notably, this degeneracy in the entanglement spectrum is not a manifestation of the corner mode, but a consequence of the projective symmetry due to the anomalous flux. Our result suggests that the entanglement features of HOTSC also reveal unique properties of topological flux responses, and that the projective symmetry in the anomalous flux state can be observed from the entanglement Hamiltonian. This paves the way for a new promising route for exploring HOTSC in numerical simulations.

\section{Projective symmetry in the superconducting flux of 2D HOTSC}
\label{sec:2d}

This section investigates the topological feature of $\pi$ flux inside a 2D higher-order topological superconductor (HOTSC) protected by $C_4$ and fermion parity symmetry. 
The motivation comes from the expectation that for a topological quantum phase protected by symmetry $G$, one can detect its topological feature by observing the anomalous symmetry structure inside the symmetry defect (e.g., the gauge flux for $G$ symmetry)\cite{ryu2015interacting,levin2012braiding,chen2017symmetry}. For instance, in a 2D $p+ip$ superconductor with fermion parity symmetry, the superconducting vortex contains a Majorana zero mode\cite{read2009non}. In 3D $\mathcal{T}$-invariant topological superconductor, the $\pi$ flux line carries a 1D chiral Majorana mode with $c=1/2$ central charge\cite{qi2009time}. For a general $G$ symmetry-protected topological phase, once we gauge the symmetry $G$, the symmetry flux either carries a fractional quantum number of $G$ or contains anomalous gapless modes. This aspect provides an alternative way to visualize the underlying quantum structure of the symmetry-protected topological phases. In addition, exploring symmetry flux offers a feasible way to detect topological responses via numerical simulations or experimental measurements.

\begin{figure}[h]
    \centering
    \includegraphics[width=2.1in]{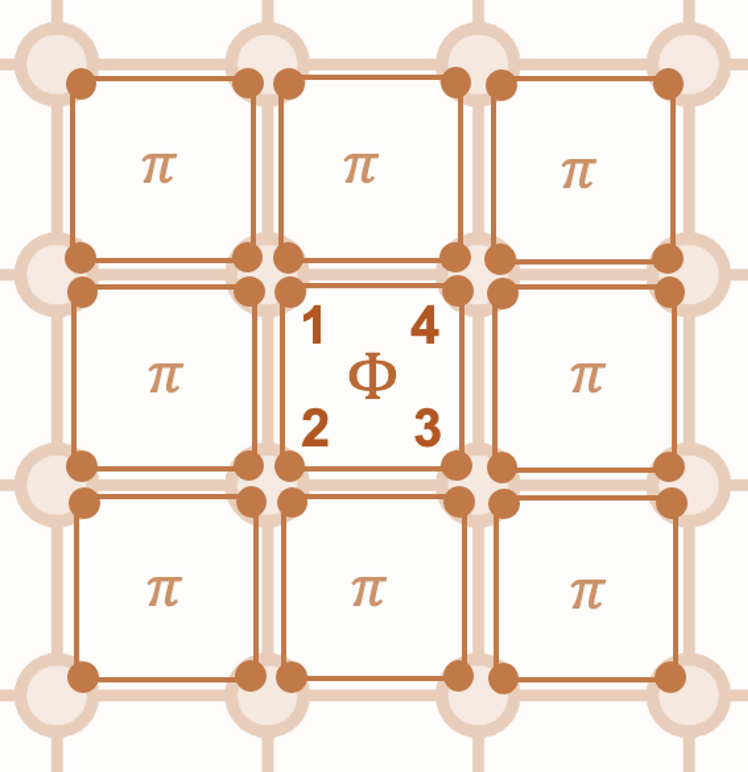}
    \caption{The HOTSC on the square lattice with four Majoranas per site. In the zero-correlation length limit, the four Majoranas on the four corners of the square hybridize within the plaquette (solid lines). The ground state Hamiltonian contains $\pi$ flux per plaquette. By inserting a superconducting $\pi$ flux in the center, the central plaquette becomes $\Phi=0$.}
        \label{2d}
\end{figure}

To set the stage, we begin with the 2D higher-order TSC proposed in Ref.~\cite{wang2018weak}. The model contains four Majoranas (two complex fermions) living inside each unit cell on a square lattice as shown schematically in Fig.~\ref{2d}. The four Majoranas at the corners of the square mainly tunnel with its nearest neighbor within the plaquette with $\pi$ flux per square. The resultant superconducting state is fully gapped inside the bulk and on the smooth edges, while the corner intersecting two boundaries contains a Majorana zero mode(MZM).

Our model contains a plaquette-centered $C_4$ rotation symmetry in addition to the fermion parity conservation $P$. The non-interacting Hamiltonian in the Majorana basis is,
\begin{align}
    &H=\eta^T (t+\cos(k_x))\Gamma^3+(t+\cos(k_y))\Gamma^1 \nonumber\\
    &+\sin(k_x)\Gamma^4+\sin(k_y)\Gamma^2) \eta\nonumber\\
    &\Gamma^1=-\sigma^y \otimes \tau^x,\Gamma^2=\sigma^y \otimes \tau^y,\Gamma^3=\sigma^y \otimes \tau^z,\nonumber\\
   & \Gamma^4=\sigma^x \otimes I.
    \label{eq1}
\end{align}
With $\eta^T=(\eta_1,\eta_2,\eta_3,\eta_4)$ being the four Majoranas on each site. $t$ is the strength for intra-site Majorana coupling. For $|t|<1$, the system is in the HOTSC phase. When $t=0$, the model returns to the zero-correlation length limit that the four Majoranas on the four corners of each square only hybridize within the plaquette.
The $C_4$ rotation acts on $\eta$ as,
\begin{align}
C_4:
   \begin{pmatrix}
0 & \tau^z \\
\tau^x & 0 
\end{pmatrix}
\end{align}

This model was widely explored in various works of literature for weak and strongly interacting systems\cite{wang2018weak,benalcazar2017quantized,benalcazar2019quantization}. The $\pi$ flux inside each plaquette is essential to acquire a fully gapped bulk spectrum and the resultant $C_4$ symmetry has the structure with $(C_4)^4=-1$ due to the $\pi$ flux.
In particular, the Majorana zero mode localized at the corner is robust against any interaction provided the $C_4$ symmetry is unbroken and the SC bulk is gapped. It was pointed out that if one develops a symmetry defect of the $C_4$ symmetry, namely, a $\pi/2$ disclination by removing a quadrant and reconnecting the disclination branch cut, there exists a Majorana zero mode localized at the disclination core\cite{you2018higher,li2020fractional}.

Now we consider the symmetry flux of the fermion parity $P$ by inserting an additional $\pi$ flux to the center of the plaquette as Fig.~\ref{2d}. As the pairing Hamiltonian already contains a $\pi$ flux in each plaquette, the additional flux insertion erases it and makes the central plaquette flux free $\Phi=0$. In the limit $t=0$ in Eq.\ref{eq1}, the flux insertion only changes the four Majorana coupling at the central plaquette, leaving a two-fold degeneracy at the center. Here and after, we will demonstrate that this degeneracy is robust against any interaction or coupling due to the projective symmetry.

To demonstrate, we begin with the particular case with $t=0$, but our demonstration is adaptive to more general circumstances, as we will elaborate on later. In this zero correlation-length limit, the four Majoranas at the corners of the central plaquette coupled like a 1D ring with four Majoranas on four sites. The ground state containing $\pi$ flux inside the plaquette imposes an anti-periodic boundary condition for the 1D ring. The plaquette-centered $C_4$ rotation symmetry performs as a `translation symmetry' that permutes between the four Majoranas on the ring. Due to the $\pi$ flux at the center, the $C_4$ symmetry permutes the four Majorana as,
\begin{align}
  & C_4(\pi): \gamma_1 \rightarrow \gamma_2,\gamma_2 \rightarrow \gamma_3,\gamma_3 \rightarrow \gamma_4,\gamma_4 \rightarrow -\gamma_1   
\end{align}
So $(C_4)^4=-1$. If we perform this rotation to the fermion-parity symmetry operator $P=\eta_1 \eta_2 \eta_3 \eta_4$,
\begin{align}
  & C_4(\pi)P C^{-1}_4(\pi)  = -\eta_2 \eta_3 \eta_4 \eta_1=\eta_1 \eta_2 \eta_3 \eta_4=P
\end{align}
The fermion parity and $C_4$ rotation commute.

However, once we insert an additional flux in the center, the `net flux plaquette' can be viewed as a ring of four Majoranas with periodic boundary conditions. The $C_4$ symmetry is now defined as,
\begin{align}
  & C_4(0): \gamma_1 \rightarrow \gamma_2,\gamma_2 \rightarrow \gamma_3,\gamma_3 \rightarrow \gamma_4,\gamma_4 \rightarrow \gamma_1   
\end{align}
Inside the superconducting flux, the $C_4$ symmetry anti-commutes with the fermion parity operator.
\begin{align}
  & C_4(0)P C^{-1}_4(0)  = \eta_2 \eta_3 \eta_4 \eta_1=-P
\end{align}
This anti-commutation relation guarantees an additional zero mode at the flux center.

To this end, we demonstrate that adding a $\pi$ flux at the rotation center engenders a projective symmetry such that the fermion parity and $C_4$ symmetry anti-commute in the flux center. As a result, the flux would trap two degenerate modes with different fermion parity connected by a $C_4$ rotation operation. We can label these degenerate modes as the even(odd) fermion parity state $|\Psi\rangle_0^a (|\Psi\rangle_0^b)$ that is related by $C_4$.
\begin{align}
  & C_4|\Psi\rangle_0^a = |\Psi\rangle_0^b
\end{align}

Our demonstration above is based on the zero-correlation limit in the absence of interaction. Now and after, we will extend this argument to a more general case with additional symmetry-preserving coupling and interaction. If the correlation length is finite, the four Majoranas in the central plaquette with net flux would unavoidably be coupled with the rest of the system. As long as there is no gap closing in the bulk, the flux state wave function $|\Psi\rangle$ can be connected to the aforementioned zero-correlation length limit wave function by a set of finite depth local unitary circuit $U$\cite{chen2010local} that commutes with the $C_4$ symmetry and fermion parity. 
\begin{align}
  & U|\Psi\rangle_0^a = |\Psi\rangle^a,U|\Psi\rangle_0^b = |\Psi\rangle^b,
\end{align}
As the unitary operator commutes with all symmetries, the new degenerate states $|\Psi\rangle^a,|\Psi\rangle^b$ carrying different fermion parity are still related by a $C_4$ rotation. This concludes that the $C_4$ symmetry and fermion parity always anti-commute inside the flux regardless of the correlation length or additional interaction. Thus, due to the projective symmetry structure inside the flux, there is no consistent way to hybridize or lift the degeneracy that preserves both rotation and fermion parity.

To summarize, we elucidate that the insertion of $\pi$ flux in a higher-order topological superconductor engenders a projective symmetry between $C_4$ and fermion parity so the resultant flux state contains a localized zero mode. In particular, inserting a gauge flux is expected to engender a projective symmetry of $C_{2n} P$. Our argument can be generalized to other 2D higher-order topological superconductors with $C_{2n}$ symmetries\cite{zhang2022bulk}.

\subsection{Emergent supersymmetry (SUSY) inside the flux}

There had been growing interest in realizing supersymmetry in solid-state systems, which is a highly appealing concept from particle physics, relating bosonic and fermionic modes. In this section, we establish a general theorem that all 2D HOTSC exhibit N=2 SUSY algebra inside the superconducting flux. As is demonstrated in Sec.~\ref{sec:2d}, adding a flux to HOTSC creates a projective symmetry between fermion parity and rotation. We will show that all flux states in HOTSC have an underlying N = 2 supersymmetry and explicitly construct the generator of the supersymmetry\cite{hsieh2016all}. 

To set the stage, we first shift all the eigenvalues of the Hamiltonian by a constant so that they are all non-negative. Then we define the following fermionic, non-Hermitian operator based on $C_4$ and P,
\begin{align}
 Q=\sqrt{\frac{H}{2}}C_4(1+P),~Q^{\dagger}=\sqrt{\frac{H}{2}}(1+P)(C_4)^{-1}. 
 \label{susy}
\end{align}
$Q^{\dagger}$ commutes with the HOTSC Hamiltonian $[H,Q^{\dagger}] = 0$. Most importantly, due to the projective symmetry inside the flux, the $Q, Q^{\dagger}$ obeys the algebra:
\begin{align}
(Q)^2=0, (Q^{\dagger})^2=0,~ Q Q^{\dagger}+Q^{\dagger} Q=2H
\end{align}

Therefore, $Q$ is the generator of an $N = 2$ supersymmetry. Such supersymmetry naturally explains the degenerate modes inside the flux. By adding a flux to HOTSC, all energy levels are doubly degenerate, and the corresponding eigenstates can be chosen as fermion parity eigenstates with different parity.  $Q,Q^{\dagger}$ operators, assisted by spatial rotation, play a role of exchanging between these fermion parity sectors. Notably, while a wide range of emergent supersymmetry in condensed matter typically requires fine-tuned Hamiltonians or critical points\cite{grover2014emergent}, the $N=2$ SUSY in HOTSC flux is 
guaranteed by the projective symmetry of $C_4P$, and thus robust against perturbations.

\subsection{Detecting flux responses from the entanglement Hamiltonian}

In this section, we show that the topological flux response in the higher-order topological superconductor can be assessed using a quantum information perspective by examining the entanglement Hamiltonian. This method not only enables the detection of the higher-order topological superconductor numerically but also suggests that the hidden topological structure, including the topological flux response, can be understood through an entanglement viewpoint.

\begin{figure}[h]
    \centering
    \includegraphics[width=3.5in]{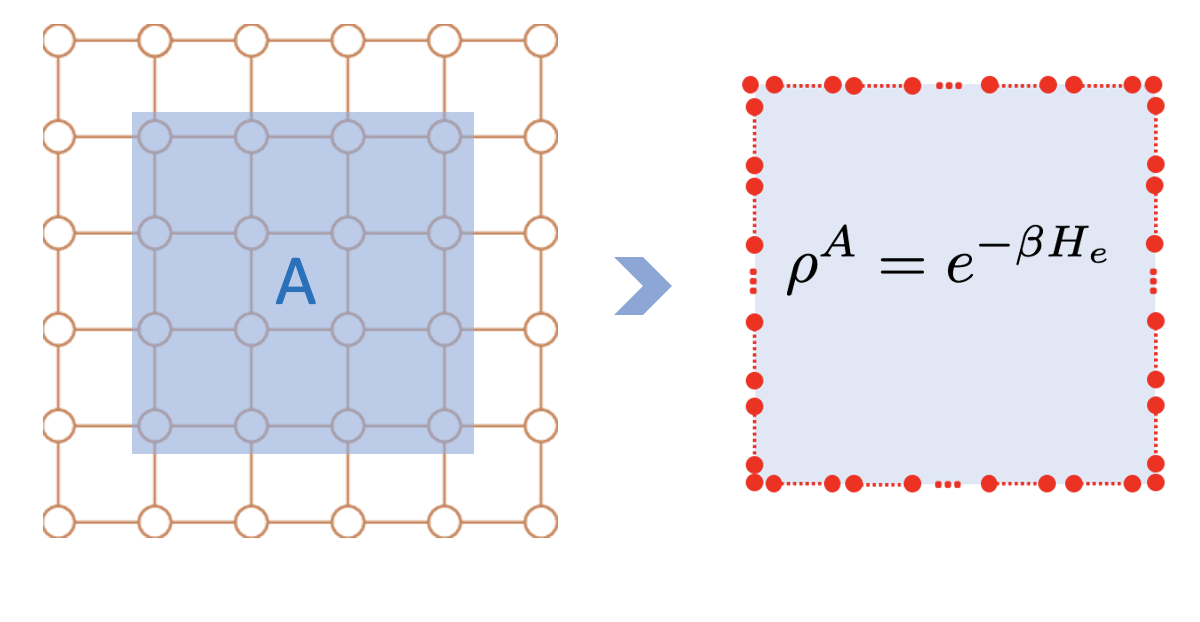}
    \caption{Tracing out the central block of the wavefunction arrives at the reduced density matrix $\rho^A$ that resembles a 1D Majorana chain along the square cut. Each quadrant of  $\rho^A$ contains odd number of Majoranas.}
    \label{entangle}
\end{figure}
To begin with, we trace out the system's central block with a size larger than the correlation length but still finite compared to the thermal dynamical limit. The resultant reduced density matrix $\rho^A=e^{-\beta H_{e}}$ can be viewed as a partition function of an entanglement Hamiltonian $H_{e}$ that resembles a 1D `square frame' along with the cut in Fig.~\ref{entangle}. Such a cut contains four corners, with each quadrant carrying an odd number of Majoranas. The $C_4$ rotation operator performs as a translation operator $T_{L/4}$ on the 1D entanglement Hamiltonian that shifts the fermion by a quarter of the cut size. In the thermal dynamical limit, the four Majoranas at the corners of the `square frame' generate four Majorana zero modes in the entanglement Hamiltonian $H_{e}$. However, these zero modes could be hybridized with finite-size cuts. Suppose we choose the length of the 1D entanglement Hamiltonian being $L$, the coupling strength between the four corner Majoranas in the entanglement spectrum scales as $e^{-\xi/L}$ so the Majorana zero-mode hybridization is inevitable for a finite-size system.
In addition, the correspondence between the `ground state of the entanglement Hamiltonian' and the wave function correlation
cannot be taken too literally. 
Since the reduced density matrix is the partition function of the entanglement Hamiltonian(EH) at finite temperatures, the high energy modes in the entanglement spectrum(ES) also contribute to the entangled features of the ground state. In particular, the low-lying states of the ES may undergo a phase transition while the bulk phase remains unchanged~\cite{chandran2014universal}. 

In terms of the ground state wave function, the central block in the reduced density matrix (with an odd number of plaquettes) contains a total $\pi$ flux. The entanglement Hamiltonian defined on the ring with $\pi$ flux inside has an anti-periodic boundary condition. The resultant $C_4$ symmetry operator of the 1D entanglement Hamiltonian can be defined as,
\begin{align}
  & C_4(\pi): \gamma_i \rightarrow \gamma_{N+i},\gamma_{N+i} \rightarrow \gamma_{2N+i},\gamma_{2N+i} \rightarrow \gamma_{3N+i},\nonumber\\
  &\gamma_{3N+i} \rightarrow -\gamma_i   
\end{align}
Here $i$ labels the Majorana on each quadrant, with 4N being the total number of Majoranas in the effective 1D entanglement Hamiltonian $H_e$.
It is not hard to check that the 
$C_4$ rotation and fermion parity commute for the entanglement Hamiltonian $H_e$. This also agrees with the fact that the entanglement Hamiltonian can have a fully gapped spectrum for a finite-size system.

To visualize the projective symmetry and zero modes inside the $\pi$ superconducting flux from the entanglement Hamiltonian, we look into the wave function with an additional $\pi$ flux in the center (so the central plaquette has net flux) and trace out the center block to get the entanglement Hamiltonian $H_e^{\text{flux}}$. Due to the additional flux insertion, the 1D entanglement Hamiltonian $H_e^{\text{flux}}$ has net flux inside the ring with periodic boundary conditions. The resultant $C_4$ symmetry operator of the entanglement Hamiltonian can be defined as,
\begin{align}
   C_4(0): ~&\gamma_i \rightarrow \gamma_{N+i},\gamma_{N+i} \rightarrow \gamma_{2N+i},\nonumber\\
  &\gamma_{2N+i} \rightarrow \gamma_{3N+i},~\gamma_{3N+i} \rightarrow \gamma_i   
\end{align}
$N$ is an odd number since each quadrant contains an odd number of Majoranas. After some simple algebra, we find that 
$C_4$ rotation and fermion parity anti-commute $C_4 P=-P C_4$ for the entanglement Hamiltonian $H_e^{\text{flux}}$. This indicates that these two symmetries act projectively on $H_e^{\text{flux}}$ so the full entanglement spectrum should display a robust two-fold degeneracy for all eigenstates. It is notable for emphasizing that this degeneracy has nothing to do with the corner zero modes in the original Hamiltonian that can be gapped due to the finite-size effect. The projective symmetry-enforced degeneracy can survive even for finite-size cuts and is robust against any interaction or perturbation.

\section{3D flux lines in HOTSC} \label{sec:3d}
This section extends our discussion on anomalous flux states to interacting HOTSC in 3D. We begin with the 3D HOTSC that supports chiral Majorana hinge modes proposed in Ref.~\cite{wang2018weak,langbehn2017reflection,benalcazar2017quantized,may2022interaction},
\begin{align}
    &H=\eta^T[(1-m\cos{k_z}+\cos(k_x))\Gamma^3+(1-m\cos{k_z}\nonumber\\
    &+\cos(k_y))\Gamma^1+\sin(k_x)\Gamma^4+\sin(k_y)\Gamma^2-m \sin{k_z}\Gamma^0]\eta\nonumber\\
     &\Gamma^1=-\sigma^y \otimes \tau^x,\Gamma^2=\sigma^y \otimes \tau^y,\Gamma^3=\sigma^y \otimes \tau^z,\nonumber\\
     &\Gamma^4=\sigma^x \otimes I,
     \Gamma^0=\sigma^z \otimes I.
    \label{3dmodel}
\end{align}
For $-2<m<0$, the model is in the HOTSC phase\cite{benalcazar2017electric}.
This model exhibit a special $C_4^T$ symmetry that rotates the x-y plane along with the time-reversal operation. The $C_4^T$ acts on the Majorana field $\eta$ as,
\begin{align}
   \mathcal{K}\begin{pmatrix}
0 & -\tau^z \\
\tau^x & 0 
\end{pmatrix} 
\end{align}
Notably, if we implement a dimension-reduction view by fixing the momentum $k_z$, the momentum layer with $k_z=\pi$ resembles the aforementioned 2D HOTSC with $C_4$ symmetry while the $k_z=0$ layer corresponds to the trivial one. $(C_4^T)^2=-1 $ indicates that the Hamiltonian has a $\pi$ flux penetrating each tube along the z-direction illustrated as Fig.~\ref{helical}.

Consider inserting an additional $\pi$ flux along the z-direction, the corresponding center tube contains net flux. Now and after, we will demonstrate that such flux insertion will engender a gapless 1D mode that is anomalous and cannot be manifested in pure lower-dimensional systems.

To warm up, recall our discussion in Sec.~\ref{sec:2d}, adding $\pi$ flux to 2D HOTSC give rise to a projective representation between $C_4$ and $P$. Our 3D model can be treated as layers of 2D superconductors with fixed $k_z$ momentum so that we can treat different $k_z$ layers independently. We consider two special momentum slices $k_z=0,\pi$ which resemble the $2D$ trivial superconductor and higher-order topological superconductor. Based on our discussion in Sec.~\ref{sec:2d}, it is clear that implementing a $C_4^T$ symmetry would change the fermion parity number $P(\pi)=(-1)^{n_{\pi}}$ inside the flux line that carries momentum $k_z=\pi$. Likewise, the fermion parity $P(0)=(-1)^{n_{0}}$ that carries momentum $k_z=0$ is not affected. Based on this argument, we conclude that the algebra between $C_4^T$ and fermion parity inside the flux line has the form,
\begin{align}
C_4^T P(\pi)=-P(\pi)C_4^T,
\end{align}
Unfortunately, the above argument relies on the fact that the fermion parity number in each momentum layer (with fixed $k_z$) is a well-defined quantum number. However, as our HOTSC does not require a translation symmetry, one can add disorder along the z-direction and the corresponding $k_z$ is no longer a good quantum number. Further, in the presence of strong interaction, fermions with distinct momentum $k_z$ can hybridize and interact. In this sense, $n_{\pi}$ again becomes ill-defined when the single-particle picture breaks down.

\subsection{Conflict of symmetry and quantum anomaly}
Here we provide a more detailed and systematic study of the flux state based on the symmetry anomaly argument. For concreteness, we will demonstrate that the flux line inside the HOTSC displays a quantum anomaly that can be manifested as a `conflict of symmetry.' If the 1D flux line is invariant under two independent symmetries $G_1$ and $G_2$, the theory is anomalous if gauging $G_1$ would break the symmetry of $G_2$ or vice versa\cite{cho2014conflicting}. Applying this `conflict of symmetry' criteria to our case, we will demonstrate that after gauging the fermion parity inside the flux line, one observes that a fermion parity gauge transformation inside the flux will automatically break the $C_4^T$ symmetry. This conflict of symmetry alternatively suggests that the symmetry assignment of $C_4^T$ and $P$ are incompatible with open boundaries so the corresponding 1D theory does not render a lattice realization.

\begin{figure}[h]
    \centering
    \includegraphics[width=3.5in]{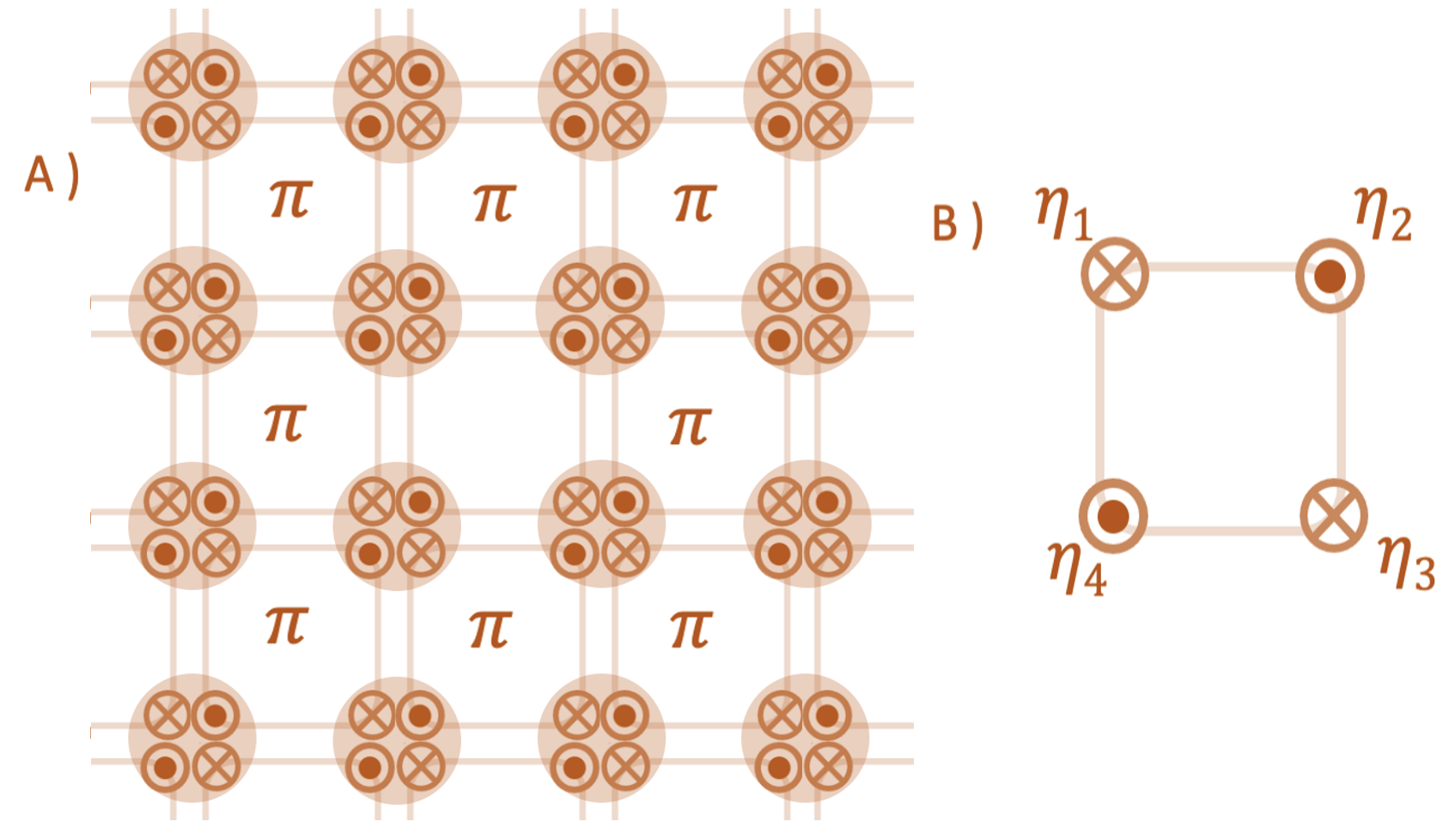}
    \caption{A) 3D HOTSC top-down view from the x-y plane. An additional $\pi$ flux penetrates the central plaquette so the total flux in the central plaquette is zero. B) Treat the z-tube as an elementary building block; it contains up-moving/down-moving chiral Majoranas living at the four hinges of the tube along the z-direction.}
    \label{helical}
\end{figure}

In the zero correlation length limit, the HOTSC model in Eq.~\ref{3dmodel} has a coupled wire construction\cite{may2022interaction}. We can decompose the complex fermions along each z-row into two up-moving and two down-moving chiral Majoranas. Treat the z-tube as an elementary building block; it contains four chiral Majoranas living at the four hinges of the tube illustrated in Fig.~\ref{helical}.
\begin{align}
H_{tube}=\eta^T (k_z) \sigma^{30} \eta
\label{1dflux}
\end{align}
We consider the general case where the four hinges along each z-tube with counter-propagating Majorana modes are coupled in a $C_4^T$ symmetric way. After inserting an additional $\pi$ flux to the central plaquette, the $C_4^T$ symmetry permutes the four components of the 1D Majorana modes in the central tube as,
\begin{align}
C_4:
   \mathcal{K}\begin{pmatrix}
0 & I \\
\tau^x & 0 
\end{pmatrix}
\end{align}
with $(C_4^T)^4=1$ provided there is net flux inside the tube center\footnote{The $\pi$ flux from the Hamiltonian and the additional $\pi$ we insert cancels.}.
The possible gapping terms for each z-tube as of Eq.~\ref{1dflux} are:
\begin{align}
&m_1=\sigma^{20},m_2=\sigma^{21},m_3=\sigma^{23},m_4=\sigma^{12},\nonumber\\
C_4^T:&m_1 \rightarrow m_2,m_2 \rightarrow m_1,\nonumber\\
& m_3 \rightarrow m_4,m_4 \rightarrow -m_3,
\end{align}
The mass terms $m_3,m_4$ are also odd under $C_2$ symmetry, so they cannot appear as a fermion bilinear mass. To make the theory compatible with $C_4^T$, we require $m_1=m_2$ and the resultant 1D flux line always remains gapless regardless of the strength of $m$.
Thus, no band mass can fully gap out the helical modes inside the flux. This in-gappable condition can be generalized to the interacting case due to the existence of an anomalous symmetry.

Now and after, we will demonstrate that the helical modes in Eq.~\ref{1dflux} are anomalous and cannot exist in pure 1D lattice models. This further suggests that the helical modes cannot be trivially gapped unless we break the $C_4^T$ symmetry. We would elaborate on this point by gauging the fermion parity symmetry inside the flux line and examining the role of $C_4^T$ under such gauge transformation.

Central to our discussion below is based on the bosonization picture of helical Majoranas in Eq.~\ref{1dflux}.
\begin{align}
&\Psi^{\dagger}_L=\eta^L_1+i\eta^L_3=e^{i\theta+i\phi+i\pi/4},\nonumber\\ &\Psi^{\dagger}_R=\eta^R_2+i\eta^R_4=e^{-i\theta+i\phi+i\pi/4}\nonumber\\
&\hat{n}=\frac{\partial_z \theta}{\pi}
\end{align}
Here $\theta,\phi$ are bosonic fields and the fermion charge density $\hat{n}$ is only defined modulo 2. \footnote{Here we add an additional $i\pi/4$ phase factor as a gauge choice that will simplify the $C_4^T$ symmetry transformation in the bosonization language}. The $C_4^T$ symmetry acts as,
\begin{align}
&\Psi^{\dagger}_L \rightarrow \Psi_R
,~\Psi^{\dagger}_R \rightarrow -i\Psi^{\dagger}_L \nonumber\\
&\theta \rightarrow \phi , ~\phi\rightarrow -\theta
\end{align}
The possible interactions that do not break $C_4^T$ are $\cos(2\theta)+\cos(2\phi)$ or their higher order descendants. Precisely, the $C^T_4$ symmetry exchange the role between particle-hole channel tunneling term $\cos(2\theta)$ and particle-particle channel pairing term $\cos(2\phi)$ by enforcing them with the same strength. These terms cannot symmetrically gap out the helical modes, so the resultant theory is either gapless or symmetry-breaking.

If we apply a gauge transformation of $P$ along the string from $-\infty$ to $z$,
\begin{align}
&G(z)=e^{i \int_{-\infty}^{z} d z' \pi n(z')}=e^{i \theta(z)}
\end{align}
Such a gauge transformation can be viewed as the fermion parity operator defined on an open string with its half-end terminated at $z$.
The $C_4^T$ symmetry transforms G(z) as,
\begin{align}
&C_4^T e^{i \theta(z)} (C_4^T)^{-1}=e^{-i \phi(z)}=-G(z)e^{-i \theta(z)-i \phi(z)}
\label{ano}
\end{align}
Such gauge transformation, equivalent to the fermion parity defined on an open chain, is not invariant under the $C_4^T$ symmetry.
Notably,$e^{-i \theta(z)-i \phi(z)}$ is a fermion operator, so the $C^T_4$ transformation creates additional fermion at the end of the fermion parity string. Such conflict of symmetry indicates that the theory cannot be placed on an open 1D chain as the fermion parity operator on the open chain is not invariant under $C^T_4$. As a result,
the helical modes inside the flux line cannot be realized in isolated 1D lattice models with the same symmetry assignment.

It is noteworthy mentioning that the conflict of symmetry was widely explored as a signature of anomalous surface states in symmetry-protected topological phases. In Ref.~\cite{cho2014conflicting,kapustin2014anomalous}, it was convinced that the conflict of the symmetries at the boundary of the SPT surfaces signals that the edge theory can never be realized as a purely lower dimensional lattice model. 
Our argument can be treated as a complement theorem signaling that the flux state inside the HOTSC also contains a gapless mode with anomalous symmetry action.



\section{Emergent HOTSC from Kitaev spin liquids}

We conclude our discussion by extending our horizon into frustrated spin systems whose emergent quasiparticle excitations exactly reproduce the topological features of HOTSC discussed in Sec.~\ref{sec:2d}. Namely, we begin with a bosonic spin model on a honeycomb lattice. Intriguingly, the low energy excitations of such a bosonic system contain emergent Majoranas coupling with an emergent $Z_2$ gauge field. The Majoranas form a superconductor reminiscent of the 2D HOTSC in Sec.~\ref{sec:2d} while the emergent flux excitation carries $N=2$ SUSY structure.

\begin{figure}[h]
    \centering
    \includegraphics[width=3.0in]{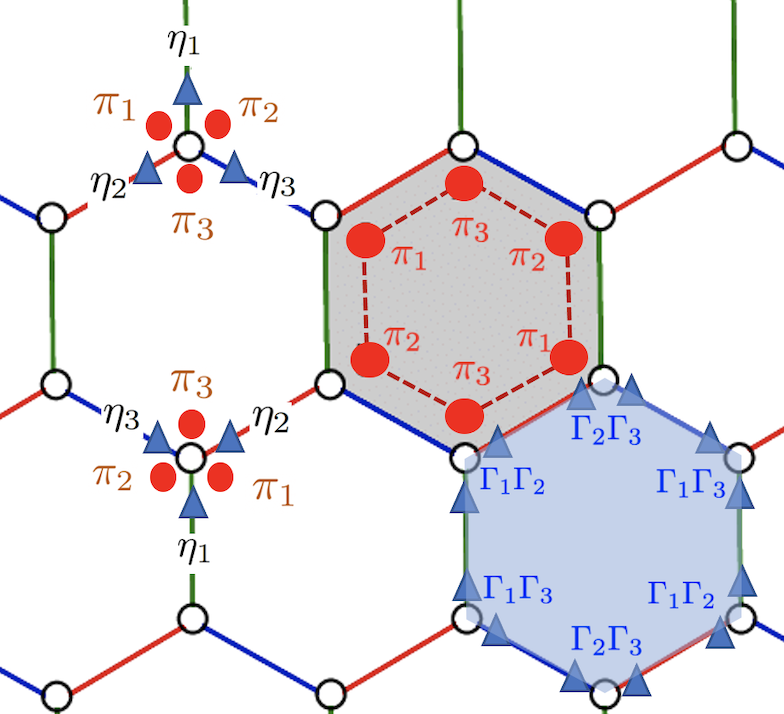}
    \caption{The spin $\frac{3}{2}$ degree of freedom and its Majorana representation on the A/B sublattice. The $\pi_i$ Majoranas are the itinerary fermions that only hybridized with their nearest neighbor within the hexagon (illustrated as the red dashed line.) The $\eta_i$ Majorana plays the role of the emergent $Z_2$ gauge potential. The Hamiltonian commutes with the flux operator defined on the blue hexagon.} 
    \label{kitaev}
\end{figure}

To continue, we focus on a specific solvable honeycomb lattice model. However, it is worth mentioning that the protocol and strategy we developed here can be applied to a wider class of lattice models, as we will elaborate on later. We begin with a spin $\frac{3}{2}$ honeycomb model with strong bond anisotropy.
\begin{widetext}
\begin{align}
&H=\sum_{i \in A, j\in B}~ [
\sum_{ij \in \text{green}}(\Gamma_1^i \Gamma_1^j-\Gamma_4^i\Gamma_1^i \Gamma_4^j\Gamma_1^j)-\sum_{ij \in \text{blue}}(\Gamma_3^i\Gamma_4^i \Gamma_3^j\Gamma_4^j+\Gamma_5^i\Gamma_3^i \Gamma_5^j\Gamma_3^j)+\sum_{ij \in \text{red}}(\Gamma_2^i \Gamma_2^j-\Gamma_5^i\Gamma_2^i \Gamma_5^j\Gamma_2^j)~]
\label{spin}
\end{align}
\end{widetext}
Here $\Gamma_a(a=1,..5)$ are the $4\times 4$ Gamma matrices with $-i\prod^{a=5}_{a=1}\Gamma_a=1$. At each A/B site on the hexagon lattice, we color three directional bonds with red/green/blue as Fig.~\ref{kitaev}. Each spin only interacts with its nearest neighbor across the red/green/blue bond, and each colored bond has two preferred spin bilinear interactions. 

Albeit the model is non-integrable, it renders an exact solvable solution inherited from the spirit of the original Kitaev model\cite{kitaev2006anyons}. In terms of parton construction, we can fermionized the spin-$3/2$ operator by introducing six Majoranas $\pi_1, \pi_2, \pi_3, \eta_1,\eta_2,\eta_3$ per site as Fig.~\ref{kitaev}. We restricted our Hilbert space with fixed onsite parity $i\pi_1\pi_2\pi_3 \eta_1\eta_2\eta_3=1$ so the six Majoranas with even parity generate a four-level system per site akin to the spin-$3/2$ degree of freedom. The spin Gamma matrices can be expressed as,
\begin{align}
&\Gamma^1=i\pi_1 \eta_1,~
\Gamma^2=i\pi_1 \eta_2,~
\Gamma^3=i\pi_1 \eta_3,~
\Gamma^4=i\pi_2 \pi_1,~\nonumber\\
&\Gamma^5=i\pi_3 \pi_1
\end{align}
So the Clifford algebra is automatically satisfied. We can express the spin operators in the Hamiltonian as,
\begin{align}
&\Gamma^1=i\pi_1 \eta_1,~
i\Gamma^4 \Gamma^1=-i\pi_2 \eta_1,~
\Gamma^2=i\pi_1 \eta_2,~\nonumber\\
&i\Gamma^5 \Gamma^2=-i\pi_3 \eta_2,~
i\Gamma^3 \Gamma^4=-i\eta_3 \pi_2,~
i\Gamma^5 \Gamma^3=i\eta_3 \pi_3
\end{align}
The model displays a special $C'_6$ symmetry that hybrid hexagon-centered $C_6$ rotation with $S_3$ spin rotation as $C'_6=C_6 \times S_3$. Under the Majorana representation, the spin rotation becomes the $S_3$ permutation between Majorana flavors,
\begin{align}
&\pi_3 \rightarrow \pi_1,~\pi_1 \rightarrow \pi_2,~\pi_2 \rightarrow  \pi_3,\nonumber\\
&\eta_3 \rightarrow \eta_2,~\eta_2 \rightarrow \eta_1,~\eta_1 \rightarrow  \eta_3,~
\label{gauge}
\end{align}

It is not hard to find a locally conserved hexagon operator illustrated in Fig~.\ref{kitaev} that commutes with all spin interactions in the Hamiltonian. This enables us to treat the $\eta_i$ fermion bilinear as the gauge potential on the link,
\begin{align}
&\exp{i\pi A_{ij \in \text{green}}}=i\eta^i_1 \eta^j_1,~~\exp{i\pi A_{ij \in \text{red}}}= i\eta^i_2 \eta^j_2,\nonumber\\
&\exp{i\pi A_{ij \in \text{blue}}}=i\eta^i_3 \eta^j_3 \label{flux}
\end{align}
$A_{ij}$ denotes the $Z_2$ gauge potential on the link between i-j sites (with i(j) belongs to the A(B) sublattice). The $Z_2$ potential on tricolored links can be written as the Majorana fermion bilinears $i\eta^i_a \eta^j_a (a=1,2,3)$ that cross between the links. As a result, the total flux in each hexagon $\oint \vec{A} d \vec{l}=\Phi$ is manifested by the product of Majorana bilinears defined in Eq.~\ref{flux} across the six links along the hexagon loop, which returns to the hexagon operator in Fig.~\ref{kitaev}.

Our argument makes it clear that, under the Majorana representation of spin operators, the $\eta$ fermion plays a role as the $Z_2$ gauge potential akin to the original Kitaev model. Likewise, the $\pi_a(a=1,2,3)$ fermion can be treated as the itinerary Majoranas that hop between nearest sites with minimal coupling to the gauge potential $A_{ij}$ on the link. To manifest, we can decompose the spin interactions as
\begin{align}
&\sum_{ij \in \text{green}}\Gamma_1^i \Gamma_1^j
=-\sum_{ij \in \text{green}}\pi_1^i \eta_1^i \pi_1^j \eta_1^j \nonumber\\
&\sum_{ij \in \text{green}}\Gamma_4^i\Gamma_1^i \Gamma_4^j\Gamma_1^j=\sum_{ij \in \text{green}}\pi_2^i \eta_1^i \pi_2^j \eta_1^j\nonumber\\
&\sum_{ij \in \text{blue}}\Gamma_3^i\Gamma_4^i \Gamma_3^j\Gamma_4^j=\sum_{ij \in \text{blue}} \pi_2^i \eta_3^i \pi_2^j \eta_3^j\nonumber\\
&\sum_{ij \in \text{blue}}\Gamma_5^i\Gamma_3^i \Gamma_5^j\Gamma_3^j
=\sum_{ij \in \text{blue}} \pi_3^i \eta_3^i \pi_3^j \eta_3^j\nonumber\\
&\sum_{ij \in \text{red}}\Gamma_2^i \Gamma_2^j
=-\sum_{ij \in \text{red}} \pi_1^i \eta_2^i \pi_1^j \eta_2^j\nonumber\\
&\sum_{ij \in \text{red}}\Gamma_5^i\Gamma_2^i \Gamma_5^j\Gamma_2^j
=\sum_{ij \in \text{red}} \pi_3^i \eta_2^i \pi_3^j \eta_2^j
\end{align}
In the Majorana representation, it is clear that all bond interactions in Equation~\ref{spin} can be treated as Majorana hopping between nearest sites, with minimal coupling to the gauge potential $A_{ij}$ represented by the $\eta$ fermion bilinears. Since the flux operators commute with the Hamiltonian, we can fix the flux sector $\Phi$ when focusing on the ground state manifold and treat $A_{ij}$ as a constant. For net flux conditions, we can simply take $A_{ij}=0$ and the permutation symmetry of the $\pi$ Majorana fermions will still hold. With $\pi$ flux patterns, any specific gauge choice of $A_{ij}$ will break the permutation symmetry of the $\eta$ Majorana fermions.

From the Majorana construction perspective, the effective Hamiltonian becomes a free Majorana model with three orbitals $\pi_1,\pi_2,\pi_3$ persite. Each orbital is only hybridized with one of the three adjacent hexagons as Fig.~\ref{kitaev}. Consequently, the fermion model is decomposed of non-overlap clusters from all hexagons. Each hexagon contains six Majoranas hybridized with their nearest neighbor that resembles higher-order topological superconductors on the honeycomb lattice\cite{zhang2022bulk}. In particular, one can easily check that the lowest energy state requires $\Phi=\pi$ flux per plaquette and the ground state is in the $\pi$-flux sector. The effective band structure for the itinerary Majoranas $\pi_1,\pi_2,\pi_3$ is reminiscent of the higher-order topological superconductor on the honeycomb lattice. Remarkably, such HOTSC may not exhibit protected corner mode for sharp corners with $2\pi/3$ angles.
Nonetheless, the topological response still holds. By creating an additional $\pi$ flux excitation in the center, the $C'_6$ symmetry and fermion parity anti-commute so the flux excitations display a projective symmetry. As both the itinerary Majorana and the $Z_2$ flux excitation originate from spin models as fractionalized excitations, the $Z_2$ flux should be treated as an intrinsic excitation rather than an external field that characterizes and probes the response. In particular, the flux excitation in this spin model contains the N=2 SUSY structure we explored in Eq.~\ref{susy}.

Finally, the construction we adopt here can be generalized to spin models on other 2D lattices. The essence relies on the fact that for any HOTSC, we can introduce a $Z_2$ gauge potential on the link and express them as a pair of 'auxiliary' Majorana fermion bilinears across the link. After onsite fermion parity projection, the resultant onsite degree of freedom becomes a hyper-spin operator, and the fermion hopping term minimal coupling to the $Z_2$ gauge potential can be written in terms of spin-bilinear interactions. Following this protocol, one can build a zoology of `Kitaev spin liquids' whose low energy excitation can constitute Majoranas with HOTSC band structure and emergent $Z_2$ gauge field. The flux excitations in these Kitaev-type models carry exotic SUSY structures with projective symmetry between spatial rotation and fermion parity.

\acknowledgments
Y.Y was supported by Gordon and Betty Moore Foundation through Grant GBMF8685 and Marie Sklodowska-Curie Actions under the new Horizon 2020.
Y.Y acknowledges informative discussions with Taylor Hughes and Rui-Xing Zhang.


%

\end{document}